\documentclass[a4paper,11pt]{article}
\usepackage{latexsym}
\usepackage{amsmath}
\usepackage{amssymb}
\usepackage{graphicx}
\usepackage{wrapfig}
\begin{document}

\baselineskip=7mm


\noindent
{\bf SAGA-HE-231-06}


\centerline{\bf Repulsive force by vector mesons and quark-hadron phase transition}

\centerline{Hiroaki Kouno}
\centerline{\it Department of Physics, Saga University, Saga 840-8502, Japan}

\baselineskip=3mm


\small
\centerline{\bf Abstract}
Using a phenomenological model with vector-type interactions, we discuss a role of repulsive force in the quark-hadron phase transition at high density. 
For realization of the quark phase at high density, strong vector coupling is needed in the hadron phase, while it is forbidden in the quark phase. 
For the quark-phase, a NJL-type model with a multi-quark interaction is investigated. 
We show that, in this model, the restoration of chiral symmetry decreases effective vector coupling and the quark phase is realized at high density, even if we have strong vector coupling at low density. 
In this model, the strong coupling in the hadron phase is induced by the chiral symmetry breaking.

~

\noindent
{\bf 1 Introduction}

To study the QCD phase transition at high density, we consider several effective models, since lattice QCD is still not feasible due to the sign problem. 
For the confinement-deconfinement (hadron-quark) transition, we often use a so-called two-phase description in which the quark and the hadron phases are described separately. 
We compare two phases by the Gibbs criteria at the same temperature ($T$) and the same chemical potential, and it is concluded that a phase which has larger pressure is realized. 
It is well-known that repulsive force between nucleons is necessary for realization of the quark-phase at high density. [1] 
In a free gas model, the ratio of the pressure ($P_{\rm Q}$) of the quark phase to that ($P_{\rm H}$) of the hadron phase becomes ${1\over{27}}$ in the high density limit, due to the relation $\mu =3\mu_{\rm Q}$, where $\mu$ and $\mu_{\rm Q}$ are the chemical potential for the baryon and the quark numbers, respectively. 
Therefore, in a free gas model, the hadron phase is realized at high density. 
The repulsive force between nucleons is needed for the realization of the quark-phase. 
In this brief report, we investigate a role of vector-type repulsive interactions in the quark-hadron phase transition at high density. 


\noindent
{\bf 2 Formalism}

For hadron phase, we use the Walecka model [2] described by the following Lagrangian. 
\begin{eqnarray}
{\cal L}_{\rm WM}
&=&\bar{\psi}\left[\gamma^\mu\{i\partial_\mu+g_\omega \omega_\mu\} -\{M-g_\sigma\sigma \}\right]\psi
\nonumber\\
&+&{1\over{2}}\partial^\mu\sigma\partial_\mu\sigma
-{1\over{4}}F_{\mu\nu}F^{\mu\nu}
-{1\over{2}}m_\sigma^2\sigma^2+{1\over{2}}m_\omega^2\omega^\mu\omega_\mu ;
~~~F_{\mu\nu}=\partial_\mu\omega_\nu -\partial_\nu\omega_\mu.~~~~~
\end{eqnarray}
The $\psi$, $\sigma$, $\omega_\mu$, $M$, $m_\sigma$, $m_\omega$, $g_\sigma$ and $g_\omega$ are the nucleon field, the $\sigma$-meson field, the $\omega$-meson field, the nucleon mass, the $\sigma$-meson mass, the $\omega$-meson mass, the $\sigma$-nucleon coupling and the $\omega$-nucleon coupling, respectively. 
The coupling $g_\sigma$ and $g_\omega$ are chosen to reproduce to the saturation properties of nuclear matter. 
We remark that the baryonic chemical potential $\mu_0$ at the normal density is equal to $M-a_1=923$MeV, where $a_1$ is the binding energy of a nucleon in nuclear matter. 

For the quark phase, we use the two flavor NJL-type model with vector interaction [3] which is described by the following Lagrangian. 
\begin{eqnarray}
{\cal L}_{\rm NJL} &=& {\bar q}( i \gamma^\mu\partial_\mu-m_0 ) q
           + {G_{\rm s}\over{2}}\Bigl( 
             (\bar q q)^2 + (\bar q i \gamma_5  \vec{\tau} q)^2 \Bigl)
             -{G_{\rm v}\over{2}}( \bar q \gamma^\mu q)^2. 
\end{eqnarray}
The $G_{\rm s}$ and $G_{\rm v}$ are the coupling constants for the scalar and the vector-type interactions, respectively. 
The $G_{\rm s}$ and the cut-off $\Lambda$ are chosen to reproduce the pion decay constant $f_\pi$ and the pion mass $m_\pi$, while we treat $G_{\rm v}$ as a free parameter. [4] 
If we introduce the auxiliary meson fields
$$\vec{\tilde{\pi}} =\bar q i \gamma_5  \vec{\tau} q,~~~\tilde{\sigma} =\bar q q,~~~\tilde{\omega}_\mu= \bar q \gamma_\mu q$$
, which correspond to the meson fields in the Walecka model, 
${\cal L}_{\rm NJL}$ is rewritten as follows. [5] 
\begin{eqnarray}
{\cal L}_{{\rm q}\pi\sigma\omega}
&=& {\bar q}( i\gamma^\mu \partial_\mu -m_0 ) q
+G_{\rm s}
\Bigl(\tilde{\sigma} {\bar q}q +
\vec{\tilde{\pi}}\cdot 
{\bar q}i\gamma_5 \vec{\tau}q   \Bigr)
-G_{\rm v}
\tilde{\omega}^\mu {\bar q}\gamma_\mu q
-{G_{\rm s}\over{2}}\left(\tilde{\sigma}^2+\tilde{\pi}^2\right)
+{G_{\rm v}\over{2}}\tilde{\omega}^\mu\tilde{\omega}_\mu
\nonumber
\end{eqnarray}

To determine which phase is realized, we should compare pressures in two phases. 
Because of the saturation properties of nuclear matter, $P_{\rm H}=0$ at the normal density. 
Therefore, for realization of the hadron phase at the normal density, 
$$P_{\rm Q}(T=0,\mu=\mu_0)<P_{\rm H}(T=0,\mu=\mu_0)=0$$
is needed. 
Since
$$\left.{\partial P_{\rm Q}\over{\partial \mu}}\right|_{T=0}={n_{\rm Q}\over{3}}\geq 0, $$
where $n_{\rm Q}$ is the quark number,  
the following inequality should be satisfied. 
$$-B\equiv P_{\rm Q}(T=0,\mu =0)^<_- P_{\rm Q}(T=0,\mu =\mu_0)<0$$
In the following, we call $B$ a "bag constant". 
Therefore, a positive bag constant is needed for the realization of the hadron phase at the normal density. 

~

\noindent
{\bf 3 Numerical results}

For numerical calculations, we use the mean field approximation. 
In Fig. 1, we show the pressures of both phases at $T=0$ as functions of $\mu^4$. 
It is seen that the hadron phase is realized at high density, if vector coupling is strong in the quark phase. 
(In this figure, we put $B=0$. 
Realization of the quark phase is less favorable for the case of a positive bag constant. )
For the realization of the quark phase at high density, strong vector coupling is needed in the hadron phase, while it is forbidden in the quark phase. 

It may be natural that, at low density, the strong vector-coupling in Eq. (1) is related to the vector coupling in Eq. (2). 
If we add the multi-quark interaction
\begin{equation}
{\cal L}_{\rm MQI}=-{G_{\rm sv}\over{2}}
\Bigl( (\bar q q)^2 + (\bar q i \gamma_5  \vec{\tau} q)^2 \Bigl)
( \bar q \gamma^\mu q)^2
\nonumber
\end{equation}
to ${\cal L}_{\rm NJL}$, 
the effective vector coupling $G_{\rm v}^*=G_{\rm v}+G_{\rm sv}\tilde{\sigma}^2$ decreases as the density increases. 
(See Fig. 2. ) [4] 
As is seen in Fig. 1, the quark-phase is realized at high density even if $G_{\rm v}^\prime \equiv G_{\rm v}^*(\mu =0)$ is large. 
With an assumption $G_{\rm vH}(\equiv {g_{\omega}^2\over{m_{\omega}^2}})\sim 3G_{\rm v}^\prime$, this result may indicate that the strong vector coupling in the hadron phase is induced by the chiral symmetry breaking. 

~

\noindent
{\bf 4 Summary}

In summary, we have investigated a role of repulsive force in the quark-hadron phase transition at high density. 
For realization of the quark phase at high density, strong vector coupling is needed in the hadron phase, while it is forbidden in the quark phase. 
The NJL-type model with a multi-quark interaction may relate the weak vector coupling in the quark phase to the strong vector coupling in the hadron phase. 
This result may indicate that the strong vector coupling in the hadron phase is induced by the chiral symmetry breaking. 

\begin{flushleft}

{\bf Acknowledgement} The author thanks T. Kunihiro, M. Yahiro, A. Nakamura, M. Matsuzaki, M. Tachibana, M. Hamada and K. Kashiwa for useful discussions. 

~

{\bf References}

[1]  See, e.g., H.~Kouno and F.~Takagi, Z. Phys. {\bf C42} (1989) 209
, and references therein. 

[2] J.~D.~Walecka, Ann. Phys. {\bf 83} (1974) 491. 

[3] See, e.g., M. Kitazawa, T. Koide, T. Kunihiro and Y. Nemoto, 
Prog. Theor. Phys. {\bf 108} (2002) 929, and references therein. 

[4] K. Kashiwa, H. Kouno, T. Sakaguchi, M. Matsuzaki and M. Yahiro, arXiv:nucl-th/0608078(SAGA-HE-229-06). 

[5] See, e.g., T. Sakaguchi, M. Matsuzaki, H. Kouno and M. Yahiro, arXiv:hep-ph/0606219, and references therein. 

\end{flushleft}

\begin{center}

\begin{center}
\begin{tabular}{cc}
\includegraphics[height=80mm,width=64mm] {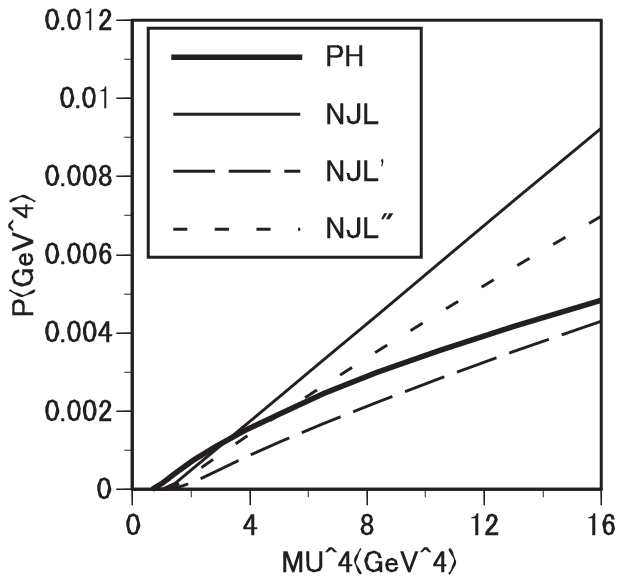} &
\includegraphics[height=80mm,width=64mm] {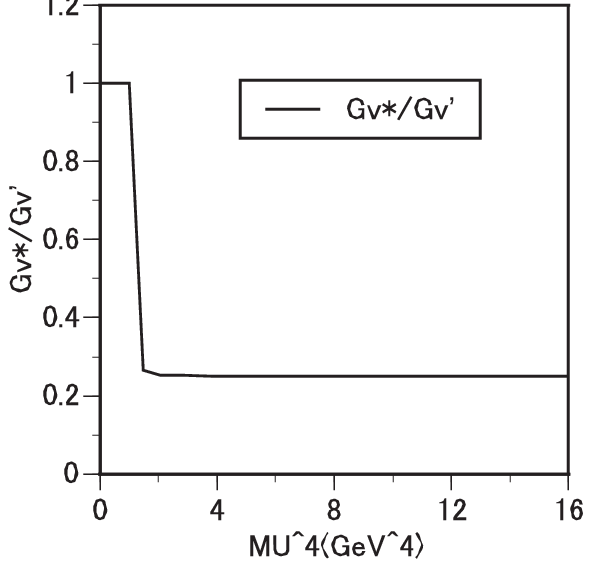}\\
$P_{\rm H}$ : Walecka model. \\
NJL : $G_{\rm v}=0$. ~~~~~~~~\\
NJL$^\prime$ : $G_{\rm v}=G_{\rm s}/1.5$. \\
NJL$^{\prime\prime}$ : ${\cal L}_{\rm NJL}+{\cal L}_{\rm MQI}$. $G_{\rm v}^\prime =G_{\rm s}/1.5$. \\
{\bf Fig.1} Pressures in the hadron and the quark phase. 
&
{\bf Fig.2} Effective vector coupling in NJL$^{\prime\prime}$. \\
~\\
~\\
~\\
~\\
\end{tabular}
\end{center}

\end{center}

\end{document}